\documentclass[]{article}
\usepackage{epsfig}

\def\FD{0.9}

\def\DP{\Delta}

\title{
Diffusion and Aggregation in an Agent Based Model of Stock Market Fluctuations
}

\author{Filippo Castiglione\\
        Center for Advanced Computer Science, University of Cologne \\
        Weyertal 80, D-50931 K\"oln, Germany \\
        {\tt filippo@zpr.uni-koeln.de}
}

\begin{document}
\maketitle
\date{}

\begin{abstract}
We describe a new model to simulate the dynamic interactions between 
market price and the decisions of two different kind of traders. 
They possess spatial mobility allowing to group together to 
form coalitions. Each coalition follows a strategy chosen from a 
proportional voting ``dominated'' by a leader's decision. 
The interplay of both kind of agents gives rise to complex price dynamics that 
is consistent with the main stylized facts of financial time series.
\\
The present model incorporates many features of other known models and is meant
to be the first step toward the construction of an agent-based
model that uses more realistic markets rules, strategies, and information structures.

\smallskip\noindent
{\it keywords}: Financial market, agents-based models, lattice gas, social influence,
collective analysis.
\end{abstract}

\section{Introduction}
In recent years many microsimulation models of financial markets are being
developed \cite{KimMark:89,LLS:94:95:97,ContBou:97,LeBaronArthurPalmer:99} 
(see \cite{StauffReview:98,LuxReview,LeBaronReview:96,LLSBook:2000,WEHWorkshop:2000} 
for a review).
All these models emphasize some aspects of the traders and their behaviour.
In \cite{ContBou:97} agents aggregate in percolation lattices to model
the tendency of traders to imitate each other; in a variant \cite{StaSorn:99} 
of this model the agents are allowed to diffuse on the lattice to induce 
autocorrelation in volatility; 
in \cite{LLS:94:95:97} the agents are equipped with a capital
that they reinvest and accumulate while in \cite{KimMark:89} the authors distinguish 
among liquidity and stocks at the current nominal price.
Moreover, some models (the first example is found in \cite{LuxMarchesi:99}) take 
into account the different tendency to follow a \emph{perceived price} for the assets 
while \cite{Farmer:2000} illustrates different strategies for value trades some of
which are much more realistic than the simple random buy/sell/inactive choice of 
\cite{ContBou:97}.
%

The present model incorporated many of these features into a single versatile model.
The main goal is to give an easy way to implement different 
key issues in modeling the stock market, to understand the relevance and 
the mutual influence of certain factors that other models have treated separately,
and to investigate the necessary and sufficient conditions determining the
factors which actually drive the empirical observed facts in real markets.

The word ``agents'' in the title of this article can be misleading due to the 
fact that traders in financial markets are also called \emph{agents}. 
Instead we refer to the idea of agents in the ``agent-based models'' in the
simulation and modeling paradigm.
The modeling and simulation of real systems consisting of agents that
cooperate with each other has emerged as an important field of
research. They are regarded as a consistent paradigm enabling an important step 
forward in empirical sciences, technology and theory \cite{Workshop2000}. 
To model the dynamics of a complex system composed of interacting entities with
their internal complex structure and dynamics has many appealing points:
self organization strategies and decentralized control,
emergent behaviour, autonomous behaviour, cooperative capacity 
and aggregation, and spatial mobility.

The purpose of the model hereafter described is to provide a relatively simple 
description of the price formation in a stock market.
The agents paradigm is most suited to accomplish this task. 
Our personal experience in modeling other complex systems rather than financial 
markets (in biology and in particular in immunology \cite{PARIMM})
tell us that complex systems need
versatile and powerful methods to be simulated with a certain degree of realism.

This paper is organized as follows: in section \ref{md} we will present the model, 
in section \ref{discussion} we will discuss its dynamics presenting the results 
of some simulations and in section \ref{conclusions} we point out some 
further developments.

\section{Model description} \label{md}

In this model we consider two kinds of agents representing what are called
\emph{chartists} and \emph{fundamentalists} 
\cite{LuxMarchesi:99,SteiglitzHonigCohen:96}. 

The first are traders and speculators whose strategy depends exclusively on the
price history. They are known also as \emph{technical traders}. 
For the time being, we model the chartists as traders who base their decision only on 
the ``current'' value of the price and not on a real historical evaluation of the ``trend''.

The other kind of traders, the fundamentalists, consider the
``fundamental'' value to determine the ``right'' price of an asset.
To do so we consider an artificially generated time series as the perceived just price
for the traded asset. 
\\
Fundamentalists are also called \emph{value-traders}.
Value-strategies are  based on perceived value, that is a model for
what something ought to be worth, which not necessarily corresponds to the actual value. 
The perceived value is inherently subjective and at a first approximation it is 
considered external information.
Basically, value strategies tend to buy when an asset is undervalued and 
sell when it is over-valued.

This distinction between the two kind of traders is mostly reflected in the
different way they decide, at each time step, to buy or to sell a stock.
This will be treated in section \ref{trading}.

\noindent
Each agent $i$ is represented by some attributes.
The order at time $t$ $x^{(i)}_t \in\{+1,0,-1\}$, indicates respectively the decision to buy, 
to sell or to stay inactive. 
%
The capital $c^{(i)}_t$ of agent $i$ at time $t$ is the current amount of money or 
credits. Each agent starts with a certain capital of money with which (s)he can
buy stocks. The decision to buy/sell is then constrained by the 
availability of stocks to be sold or funds to buy other stocks. 
Agents reinvest their profit and accumulate capital.
The number of owned stocks of agent $i$ is indicated by $n^{(i)}_t$. 
At each time step, the owned stocks account for the actual wealth of the trader 
with the current nominal price of the asset $p_t$.

Along with the agents we need to define how an asset is represented in our
artificial stock market, that is, how the price is determined.
The current price $p_t$ is computed every time step by a single market maker.
While the traders observe the actual price or the perceived price and 
submit orders $x^{(i)}_t$, the market maker fills all orders at the new
price $p_{t+1}$ according to a certain increasing function of the excess demand
$p_{t+1}\propto \zeta(p_t,D_t)$
where $D_t=\sum_i x^{(i)}_t$ denotes the total excess demand a time $t$.
The form of the function $\zeta$ is usually very complicated and also takes into account
exogenous factors (for example the market maker wants to adjust the price so as to 
eliminate eventual profit opportunities). 
For sake of simplicity we set the new price proportional to the
square root of total supply and demand (as in \cite{Zhang:99})
\begin{equation}
 \DP_{t} =  p_{t+1} - p_{t} \propto {\rm sign}(D_t) \sqrt{|D_t|}
\label{pricing}
\end{equation}
\noindent
On the other side, to model the fundamentalist behaviour, we need to define the 
perceived value $f_t$ of the asset.
This include an exogenous source of information hitting the market,
thus $f_t$ can be arbitrarily modeled as an exogenous stochastic process 
(\cite{LuxMarchesi:99}) with the relative 
changes drawn from a Gaussian with zero mean and standard deviation $\sigma_\epsilon$:
$\log(f_{t+1}) - \log(f_t) = \epsilon_t$
where $\epsilon_t \sim {\cal N}(0,\sigma^2_\epsilon)$. 
Note that the fundamental value $f_t$ is perceived 
at the same time and in the same way by all the fundamentalist traders. 
In \cite{Farmer:2000} also the case in which there are diverse perceived values is 
studied.
\\
At each time step both $p_t$ and $f_t$ are updated as specified above. 
In particular $p_t$ is updated with eq(\ref{pricing}) where the strategy $x^{(i)}_t$ 
for each agent $i$ is determined by the trading rule described in the next section.
The time corresponding to a single iteration should be thought as the time scale
on which the fastest traders observe and react to price, e.g., a minute to a day.
In section \ref{discussion} we will determine approximatively the time scale of the 
model just looking at the outcome of a simulation.

\subsection{Trading strategy}\label{trading}
This section describes the logic which rules the trading activity.
At each time step each agent decides, independently from the others, if to buy,
to sell or not to trade.
The probability to stay inactive is $1-\alpha$ where $\alpha$ is the 
activity parameter as in \cite{ContBou:97}.
In the case they choose to trade, a different decision path is followed by the
two types of traders.

\paragraph{Fundamentalist strategy}
Fundamentalists tend to follow the ``right'' price $f_t$. 
At each time step the decision to buy or sell proceeds as follows:
\vskip0.2cm\noindent
(a) If they own stocks then they check if it is appropriate to sell:
if $p_t > f_t$ then they sell, otherwise they buy another stock. 
If they cannot buy because no money is left, then they stay inactive.
\vskip0.2cm\noindent
(b) In the case they do not own any stock,
they check if it is appropriate to buy, that is, if $p_t<f_t$. 
If $p_t<f_t$ and if they have enough money, they buy, otherwise they stay inactive.

\vskip0.2cm\noindent
Note that this is similar to the ``order-based value strategy'' in \cite{Farmer:2000} 
that buys \emph{until prices match values}. That strategy is consider unrealistic 
because it is not bounded by any risk constraint. In the present rule, instead,
the upper/lower bound is given respectively by the limited amount of capital of 
the agents and by the number of stocks they own.

\paragraph{Chartist strategy}
Chartists follow the market price $p_t$. They are not influenced by any external 
source of information. At each time step they decide as follows:
\vskip0.2cm\noindent
(a) If they own stocks, then they ask if it is appropriate to sell, that is, if 
the average price they paid for all the stocks they have bought up to then is lower 
than $p_t$. 
For example if one has invested $p_{t'} + p_{t''}$ (with $t',t''<t$) in two stocks
and if $p_t > (p_{t'} + p_{t''})/2$ then selling one of those stock will be a 
profit opportunity. 
If this does not hold and (s)he has money, (s)he will buy another stock.

\vskip0.2cm\noindent
(b) In the case they do not own stocks, they simply flip a coin to decide 
whether to buy or to do nothing.

\vskip0.2cm\noindent
Case (b) gives a larger degree of randomness to the behaviour of the chartists.

\subsection{Collective formation and diffusion}\label{aggregate}
At this stage of formulation the agents follow independent strategies.
Thus, the excess demand $D_t$ is the sum of independent identically 
distributed random variables and for the central limit theorem determines 
a Gaussian distribution of the histogram of log-return which is different 
from what is observed in the real markets \cite{Mantegna:95}.

To obtain a deviation from normality we need to take into account
\emph{herding} behaviour that has already been demonstrated to determine (at
least in part) the fat tail property of the distribution of returns
\cite{ContBou:97,StaSorn:99,ChowStauf:99,StaufferYan:2000}.
\\
Instead of determining a priori the clusters distribution as in percolation 
models we just allow the agents to diffuse on the grid and 
to aggregate inside each single lattice site 
(in \cite{StaSorn:99} self-organized percolation is used to have percolating 
clusters without setting the system to the critical point). 

To give spatial mobility to the agents we model the market
as a \emph{lattice gas}.
Agents are placed randomly on a regular hexagonal (honeycomb) 
two-dimensional lattice $L\times L$ (six links at $60^\circ$) with 
toroidal boundary conditions.
Each site of the lattice may potentially contain all the agents.
At each time step, agents diffuse uniformly to a neighboring site.
The diffusion determines a re-shuffling of agents inside each single 
lattice site and changes the impact on the price fluctuations.

Now, the decision of each agent to buy, sell or stay inactive is determined in 
two phases: first, each agent makes his choice according to his actual situation
and potential benefit from the activity as specified by the trading strategy in section
\ref{trading}, then, in the second phase, the single agent's
strategy is aggregated to a simple collective strategy.
In this last phase the agent's decisions $x^{(i)}_t$ are weighted by the 
\emph{influence strength} $s^{(i)}$ in a kind of proportional vote described below. 
\\
In the spirit of Nowak et al. \cite{Nowak:90} (see also \cite{ChowStauf:99})
the decision of each agent affects and gets affected by other agents
on the basis of its influence strength.
The influence strength of agent $i$ is represented as a real number $s^{(i)}$. 
If we assign a much larger influence to some traders among the totality 
(call them \emph{leaders}), the dynamics of the model will be dependent on how many 
leaders are present.
Leaders are chosen uniformly among the two classes
of traders, chartists and fundamentalists.
We represent a group of traders by those agents contained in same lattice site.
So, each group of traders forms a collective system.
We also want a group to be dominated by a \emph{leader} so that
we can set the number of leaders equal to $L^2$ given that 
$N\gg L^2$ where $N$ indicates the total number of agents.

The influence among agents found in the same lattice site works as follows:
each agent $i$ imposes his/her strategy according to the influence strength $s^{(i)}$. 
For each $x^{(i)}_t\in\{-1,0,+1\}$ determined by the trading strategy discussed above, 
we sum the influence strength of each agent
that is following strategy $x$ and we normalize it to the total so to get 
values between zero and one to be interpreted as probabilities ($\sum_x pr[x]=1$):
\begin{equation}
 pr[x] = {
              \sum_{i:x^{(i)}=x} s^{(i)}
         }/{\sum_i s^{(i)}}
\label{randweel}
\end{equation}
where the index $i$ runs over all agents contained in that lattice site.
Then we update agents' strategy $x^{(i)}$ using a \emph{random wheel} 
with probability $pr[-1]$, $pr[0]$ and $pr[+1]$ given by eq(\ref{randweel}).

A collective system arises from the coherent behaviour of a group of strongly interacting 
constituents. It also has a weak external coupling.
In our case, the collective can be treated as an individual whose strategy has a 
distribution of buy/sell/inactive which is determined by the single agent's actual 
strategy $x^{(i)}$ through eq(\ref{randweel}). 
This means that if exactly one agent is a leader and he has chosen strategy 
$\hat{x}$ then $pr[\hat{x}] \gg pr[x]$ for the other choices $x$.
Thus, the fraction of agents in that collective that will follow strategy 
$\hat{x}$ is large. If more than one leader, say two, is present in 
the same site and their decision is different, then they will compete to 
determine the majority strategy. 
On average half of the traders will follow one leader and the other half will 
follow the other. And so on for the other possibilities. In general, if we set
more than one single leader in each site (on average), their competition will
destroy the effects of the herding behaviour.
This determines, as confirmed by numerical simulations, a
Gaussian distribution of returns.
\\
In conclusion we decide to set on average a single leader (or less) in each lattice 
site.

It is worth to note that even in this case the dynamics of each collective group 
is equivalent to that
of a single agent (the leader) with combined capital. Following this reasoning one
would expect a distribution of return with no fat tails at all. 
In fact the effect of the collective strategies
of the leaders with combined capital would end
up to zero when summed over the whole grid just because the uniformity of the 
collectives' dimension which is about $(N_C+N_F)/L^2$ on average. 
The results show a clear deviation from a Gaussian instead. 
Why this happens? The answer is found considering the synergy between the trading rule
of section \ref{trading} and the collective formation mechanism described in 
\ref{aggregate};
the collective groups influenced by a fundamentalist leader are coupled 
(weakly, but they do) by means of the fundamental price that is perceived by all of 
them equally. So, while the \emph{chartists' collectives}, being totally uncoordinated, 
only produce noise, it happens that fundamentalists' leaders occasionally end up with 
the same decision to buy or sell, driving a large fraction of all the traders to follow 
the same decision. 

The re-shuffling given by the diffusion of agents can also be interpreted as 
a change of preference, in the same way people decide to trust to another brokerage 
agency or bank.
Moreover, not all lattice sites will contain a leader. In these ``leader free'' sites 
no collective if formed and the behaviour of the agents is totally independent.
Another remarkable mechanism given by the diffusion is that one agent that leaves a 
group whose leadership is fundamentalist for a group whose leadership is chartist will 
end up in a \emph{behaviour's change} much like the fundamentalist-noisy switch in 
\cite{LuxMarchesi:99}.

\section{Discussion}\label{discussion}

When the number of fundamentalists is higher than that of chartists,
the price $p_t$ tracks closely the perceived value $f_t$. 
This is both trivial and unrealistic. 
In fact, evidence from real market data suggests that, while prices track values
over the very long term, large deviations are the rule rather than the
exception \cite{CampShi:88}.
The opposite situation results in too random fluctuations given the random behaviour
of the chartists with the price that does not follows the value.

So, a non trivial dynamics is obtained for $N_F \simeq N_C$. In this case 
we observe periods in which the price follows the fundamental price followed by
periods of apparently independent fluctuations.
For this purpose we set $N_F= \hat{N} L^2$ and $N_C= \hat{N} L^2$ for a 
certain $\hat{N}$. The value of this parameter is found observing that the collective
dynamics depends on its dimension. 
If the collective systems are too small we do not get any herding behaviour.

To avoid too large fluctuations of the price around the fundamental price,
we can also get rid of another parameter and set the activity $\alpha = 10/(N_F+N_C)$.

\begin{figure}[htbp] 
\begin{center}
  \leavevmode
  \epsfig{file=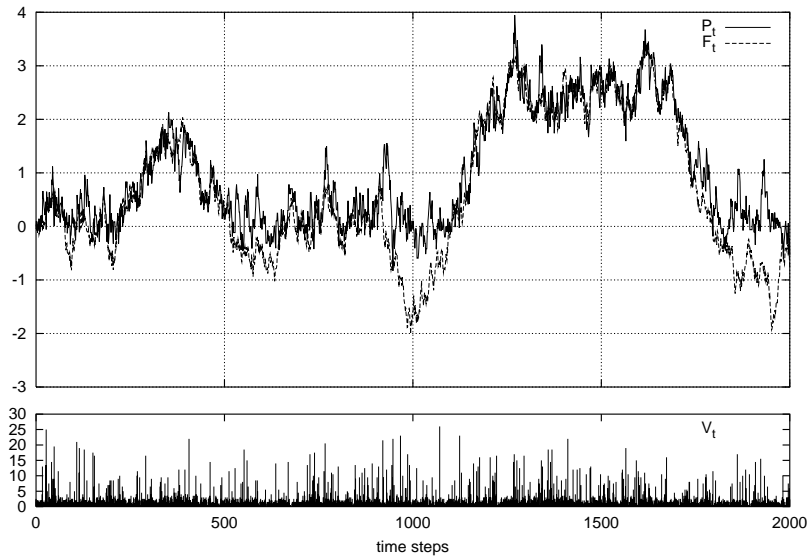,width=\FD\textwidth} 
  \caption[1]{
  Asset price $p_t$ and fundamental price $f_t$ difference with the initial value 
  as function of time for the
  first 2000 time steps of a longer simulation with 2500 agents for each type.
  In the plot at the bottom is shown the volume $V_t$ defined as the number
  of contracts between agents willing to buy and agents willing to sell.
  \label{price}
  }
  \end{center}
\end{figure}

Figure \ref{price} refers to a simulation with $L=10$ and $\hat{N}=25$ 
running for $10^6$ time steps.
The influence strength of the leader is set to 100 while the remaining agents
have influence strength one. The diffusion speed is set to $10^{-1}$.
As in \cite{Caldarelli:97} half of the agents start with one 
stock while the remaining with no stocks.
In this way we obtain a balance between an initial number of people 
willing to sell and people willing to buy that would produce an artificial initial 
bubble or valley whose amplitude depends on the number of market participants.
Besides it is worth to note that the initial amount of capital each agent is 
equipped with, induces the ability of the market price $p_t$ to follow the fundamental 
value $f_t$ when this strongly deviates from the initial value $f_0$.
To see this fact just consider the case in which the agents 
own little initial capital, then $p_t$ is limited by the global capacity 
of the agents to buy.
On the other hand, the price $p_t$ is limited from below by the amount of initial stocks
we equip the agents with; the greater it is, the larger will be a potential fall
of the price $p_t$ to follow the fundamental value $f_t$. 
These and other related questions will be investigated elsewhere.

Another consequence coming from the constraint given by the availability of funds/stocks
is that, given the trading rule described above, the activity of the agents is not 
uniform. In fact, agents may end up with the decision to stay inactive either if they 
want to sell but they do not own stocks or they want to buy but they do not have money.

\begin{figure}[htbp] 
\begin{center}
  \leavevmode
  \epsfig{file=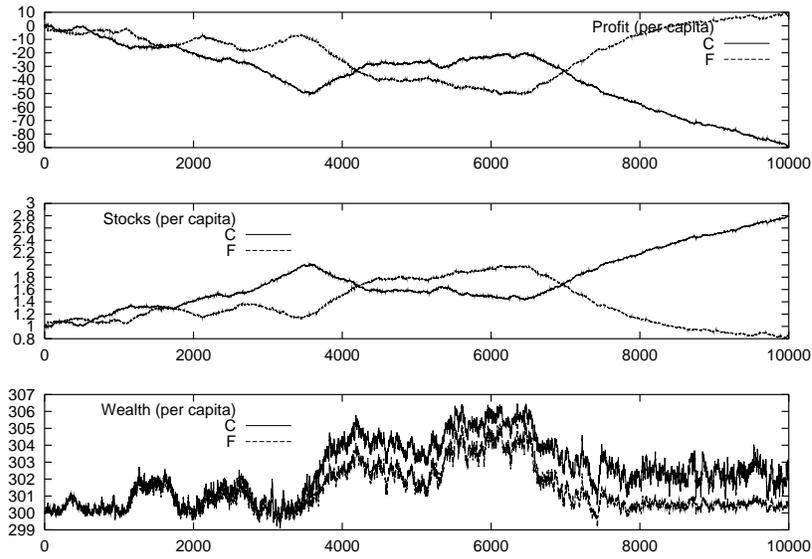,width=\FD\textwidth} 
  \caption[2]{
  The figures show only $10^4$ time steps (on the X-axis) of the much longer simulation 
  discussed in the text. 
  The upper plot shows the profit per capita for both type of agents.
  In the middle the average number of owned stocks per agent.
  At the bottom the average total wealth computed as the sum of liquidity
  and stocks at the current nominal price $p_t$. 
  The initial value of 300 (\emph{euro} or arbitrary credits) is given by 
  the initial availability 
  of capital. This plot shows that both types of agents
  have very similar patterns of earning. Here, the chartists are doing better than the 
  fundamentalists.
  \label{wealth}
  }
\end{center}
\end{figure}

Figure \ref{price} shows the price $p_t$ to follow the fundamental price 
$f_t$ apart of some large occasional deviations.
The standard deviation of the price is $\sigma_{p}= 46$ while $\sigma_{f}=44.94$ 
giving an \emph{excess volatility} of 2.3\%.
\\
In the small chart at the bottom of the same figure we show the traded volume $V_t$
computed each time step (in contrast to reality where it is computed every certain
period of time, e.g., day or week but not instantaneously).
Because in this model we do not require to match the sell/buy orders, we compute 
the volume as $V_t = \sum_i |x^{(i)}_t|$.
In fact the model does not require that each buy order should match a sell order 
because the balance is assured to be made involving a market maker outside the 
model itself.
The figure shows an increase in volume when $p_t \simeq f_t$ because, according to 
the trading rule, fundamentalists trade much more in proximity of the fundamental value.
This is opposed to periods in which they either (a) sell all stocks they own 
because the price is higher than the perceived value 
and wait to buy a new stock when the price is lower than its expected value
or (b) they have already invested all the capital and cannot buy other stocks even if 
appropriate (see also plots of wealth \ref{wealth}).

Figure \ref{wealth} bottom panel shows the average wealth per agent
computed as the sum of both liquidity and stocks for the actual nominal price of 
all agents.
The upper plot shows the average profit per capita and the middle plot the average 
number of stocks per capita.
It seems that a kind of synchronization exist between chartists and fundamentalist:
when one buys the other sells. This is given by the balance of the fundamentalists
to the escaping force of the chartists. In fact fundamentalists trade to put the market
on track again. In this view the chartists globally exercise a great indirect 
influence on the fundamentalists actions.
The same phenomena are observed in LLS's model \cite{LLS:94:95:97}
taking into account fundamentalists and trend followers 
(they call them \emph{extrapolating investors}).

\vskip0.1cm
To better investigate the final distribution of capital we run a large simulation 
involving two million agents and running for 15000 time steps.
\begin{figure}[htbp] 
\begin{center}
    \leavevmode
    \epsfig{file=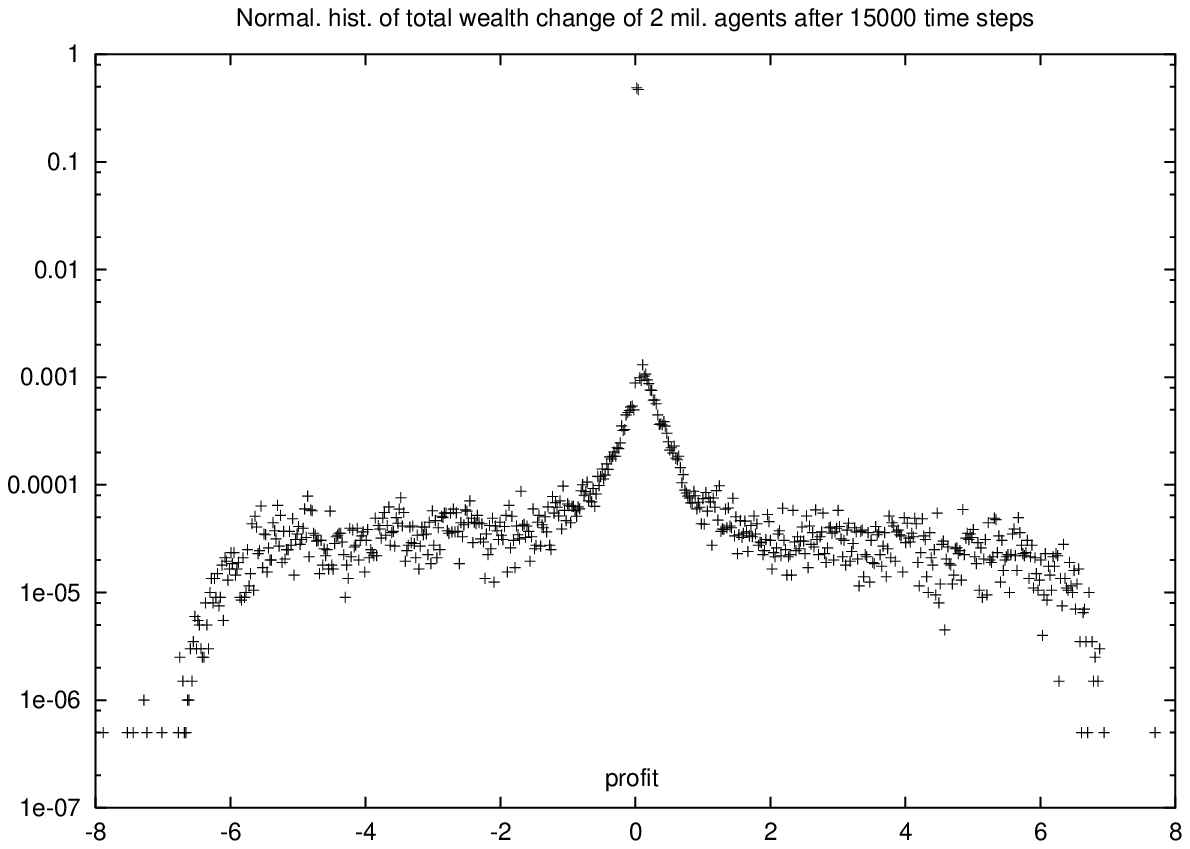,width=\FD\textwidth} 
    \caption[1]{
    On the X-axis the difference between the final accumulated capital per agent and
    the initial capital they are equipped with. A re-shaping from a uniform
    distribution (initially they have the same capital) to a power law in the
    center of the histogram is observed (Y-scale is logarithmic).
    The unit for the X-axis is given in \emph{euro} or in an arbitrary unit of money.
    \label{weahis}
    }
  \end{center}
\end{figure} 
In figure \ref{weahis} is shown the histogram of agent's wealth-change distribution.
It shows a power law in the central part with slope -1.3
and wide tails. So the central part of the distribution is consistent with a 
Pareto-like distribution \cite{Pareto:1897}. 
It is noticeable the fact that the initial uniform distribution (all agents start with
same capital) is strongly reshaped over a sufficiently long run (indeed
over shorter run the final distribution of wealth is Gaussian-like, not shown).

\vskip0.1cm\noindent
This qualitative picture of the model dynamics must be enhanced by
a more quantitative assessment that the model reproduces the
features of market price time series.

\begin{figure}[htbp] 
\begin{center}
  \leavevmode
  \epsfig{file=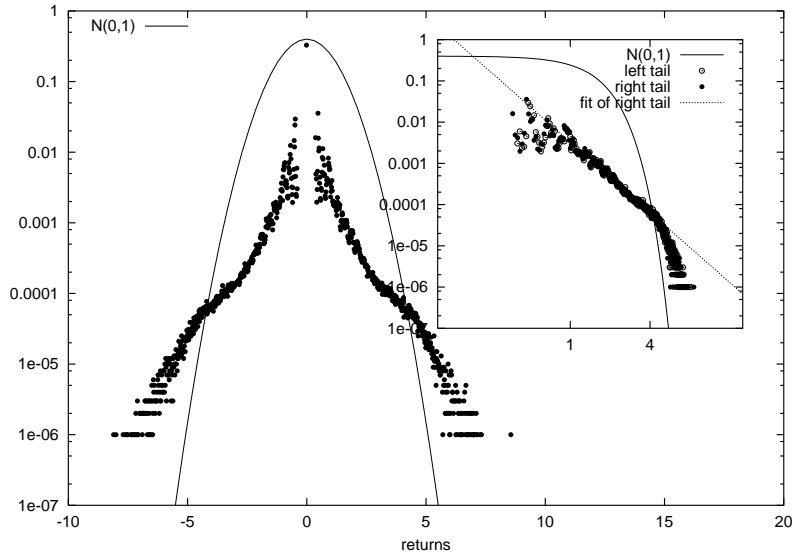,width=\FD\textwidth} 
  \caption[3]{
  Histogram of \emph{standardized} price returns $(r_t-<r_t>)/\sigma_{r_t}$ where
  $r_t = \log p_t - \log p_{t-1}$.
  The excess kurtosis is 4.58. In the log-log inset plot the fit of the central part
  of the histogram have slope $\sim 2.8$.
  \label{histo}
  }
  \end{center}
\end{figure}

Plot \ref{histo} shows the histogram of 
log-return $r_t = \log p_t - \log p_{t-1}$ 
for the same simulation of figure \ref{price} and \ref{wealth}. 
The excess kurtosis of the distribution is 4.58.
Moreover, the histogram has fat tails and power law decay $\propto x^{-\mu}$ in the
central part leading to exponent $\mu\sim 2.8$ (see inset plot) roughly consistent
with empirical studies \cite{LuxMarchesi:99,Gopi:99}.

\begin{figure}[htbp] 
\begin{center}
    \leavevmode
    \epsfig{file=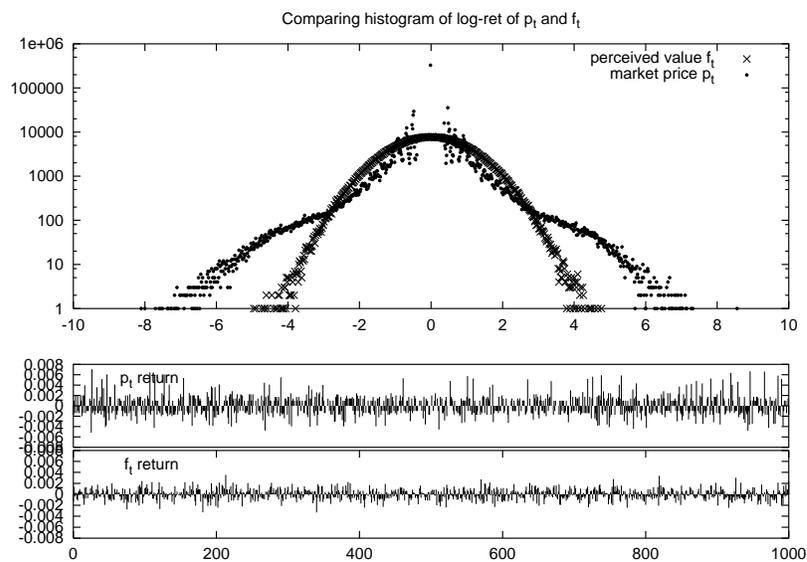,width=\FD\textwidth} 
    \caption[1]{
    Comparing the log return of $f_t$ with that of $p_t$.
    The exogenous source of information is transformed into something else 
    by the endogenous dynamics of the market participants.
    On the X-axis of the upper plot it is reported the price return while 
    on the two plots at the bottom is reported the time step.
    \label{histo2}
    }
  \end{center}
\end{figure}

Figure \ref{histo2} show the comparison of the return of the market 
price $p_t$ and those of the fundamental price $f_t$. 
The deviation is clear; the exogenous source of information
is transformed into something else by the endogenous dynamics of the market 
participant as already demonstrated in \cite{LuxMarchesi:99}. 
\vskip0.1cm\noindent
Another relevant property of market price dynamics is the absence of correlation of
return and the persistence of long range correlation of volatility 
\cite{LuxMarchesi:99,Gopi:99}.
Volatility of stock price changes is a measure of how much the market is liable
to fluctuate and can be defined in different ways. In the following we define
the volatility as the square of return $v_t=r_t^2$.

\begin{figure}[htbp] 
\begin{center}
  \leavevmode
  \epsfig{file=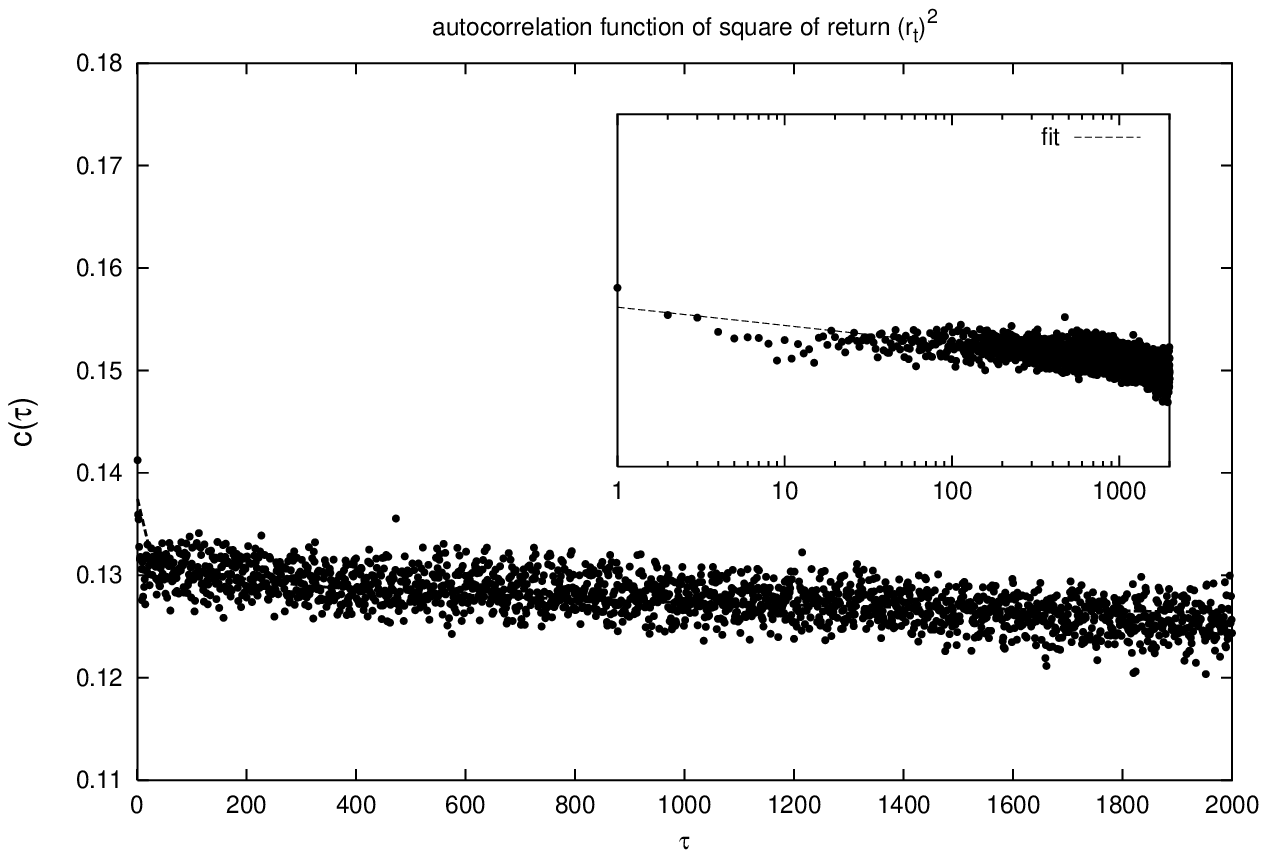,width=\FD\textwidth} 
  \caption[1]{
  Autocorrelation function of volatility of return. It shows the correlation of 
  volatility persisting for long time.
  In the log-log plot the fit has slope -0.013.
  \label{volatility}
  }
  \end{center}
\end{figure}

Figure \ref{volatility} shows the autocorrelation function of volatility defined as 
$ c(\tau)=[ <v_t v_{t+\tau}> - <v_t>^2  ]/[ <v_t^2> - <v_t>^2 ]$ .
Empirical studies show that the autocorrelation of volatility in real data follows 
a power law decay with exponent between 0.1 and 0.3 
\cite{BouchaudBook:2000,MantegnaBook:99,Ghashghaie:96}.
Instead we found a slope $\sim .013$ that is one order of magnitude less.

\begin{figure}[htbp] 
\begin{center}
    \leavevmode
    \epsfig{file=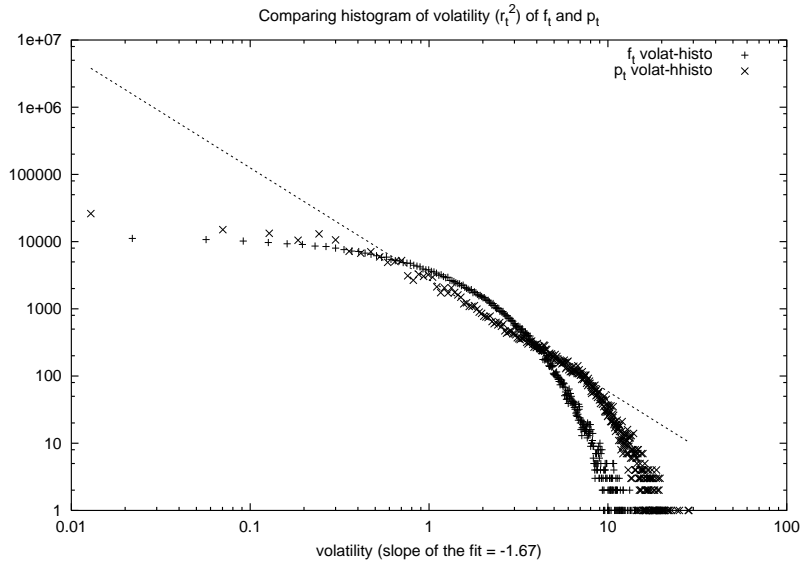,width=\FD\textwidth} 
    \caption[1]{
     Log-log plot of the histogram of volatility computed for the random
     walk $f_t$ and for the price of the asset $p_t$.
    \label{volatHist}
    }
  \end{center}
\end{figure}

Other empirical studies of large data bases \cite{LiuGopi:99} show that 
the cumulative distribution of the volatility is consistent with a power-law 
asymptotic behaviour characterized by an exponent $\sim 3$.
Figure \ref{volatHist} shows the histogram of volatility computed for
both fundamental $f_t$ and actual price $p_t$. In particular a regression in log-log
shows an exponent of -1.67 of the histogram of $r^2_t$.

In summary, the model is not yet able to perfectly match some empirical findings and
more work is required.

\begin{figure}[htbp] 
\begin{center}
    \leavevmode
    \epsfig{file=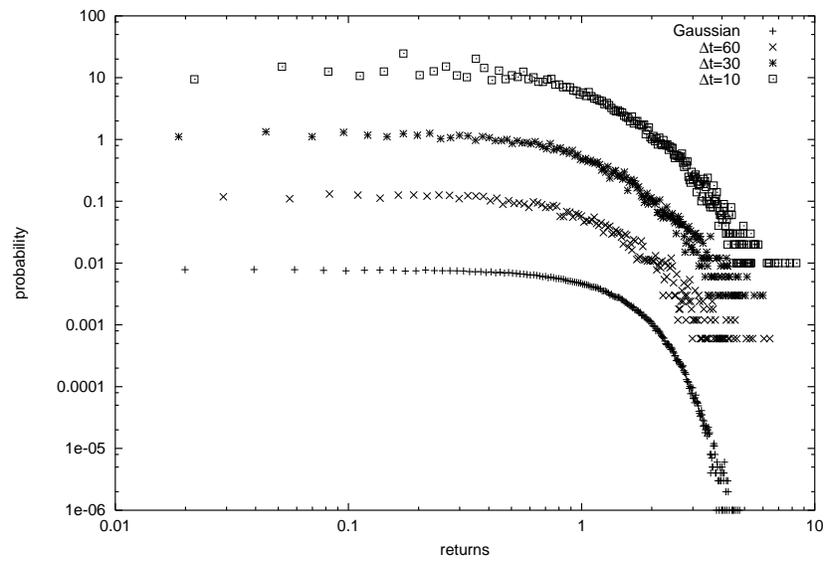,width=\FD\textwidth} 
    \caption[1]{
    Each series is vertically shifted to augment visibility.
    The crossover to a Gaussian of the histogram of the \emph{sub-series} of return
    calculated over a time scale $\Delta t=60$ time steps suggests that each time step
    roughly corresponds to half trading day in real markets.
    \label{scale}
    }
  \end{center}
\end{figure}

\vskip0.1cm\indent
To obtain a rough estimation of the time scale of the simulated market
we filtered out the time series of return $r_t$ from the complete series
at intervals of increasing size $\Delta t$. In figure \ref{scale} we show the
histogram of the filtered series for different time lags
(this is a slightly simpler procedure than the time-aggregation used in
\cite{LuxMarchesi:99}).
A crossover to a Gaussian is observed for $\Delta t \sim 60$ suggesting
that the time scale of the simulation shown in all figures but fig.~\ref{weahis}
is of the order of half day per time step.
Indeed, empirical observations tell that 
a crossover to a Gaussian is found on time scales approximatively of one month
\cite{Mantegna:95}.

There is no doubt that a more rigorous argument would be given estimating the excess 
kurtosis of the filtered series and determining the crossover when the filtered series 
with a certain $\Delta t$ have excess kurtosis about zero.
Unfortunately this would require a huge amount of data points because the kurtosis 
depends on the estimation of the fourth moment.
%
%


\section{Conclusions and future developments}\label{conclusions}

We have described a new model to reproduce the price fluctuations of a single stock in 
an artificial stock exchange whose traders are modeled as chartists and fundamentalists.
Some of them have more influence than others and represent a brokerage agency where 
people go to ask for advice.
They group together inside each lattice site to form a collective system
and to follow a common strategy according to proportional voting.
Traders are free to diffuse on a two-dimensional lattice to model the tendency 
to change opinion and to follow a different advisor.

The model is consistent with fat tails of histogram of returns and correlation
of volatility. Nevertheless much work still has to be done to better reproduce
other empirical observed facts as the volatility and wealth distribution.
\\
The structure of the model is versatile enough to allow future expansion some of 
which are discussed below. We believe that a more realistic description of the 
agents behaviour and trading may allow to get further insight in the dynamics
of price change as well as in the distribution of wealth among traders.

\vskip0.1cm\noindent
It has been made clear that the availability of capital strongly influences the 
dynamics of the model. Thus, a non uniform initial wealth distribution (some agents 
represent capitalists or large investors like banks who manage large capitals),
may determine the power law of wealth distribution.
Besides, agents should be able to buy/sell more than just one stock at a time 
according to, for example, the availability of capital and difference between perceived value and 
actual price. 
\\
Besides it is worth (and the model will easily allow it) to develop more realistic 
trading strategies like trend-following \cite{Farmer:2000} and/or
divide the action among different choices (more stocks and/or bonds with fixed 
income as in \cite{LLS:94:95:97}).

\paragraph{Acknowledgments:}
I would like to thank M.~Bernaschi, R.~Pandey and D.~Stauffer for helpful 
suggestions and constructive critics.
Fruitful discussions with Z.-F.~Huang are also acknowledged.



\end{document}